\documentclass[conference]{IEEEtran}
\IEEEoverridecommandlockouts

\usepackage{cite}
\usepackage{amsmath,amssymb,amsfonts}
\usepackage{algorithmic}
\usepackage{algorithm}
\usepackage{graphicx}
\usepackage{textcomp}
\usepackage{xcolor}
\usepackage{enumitem}
\usepackage{stfloats}
\usepackage{url}
\usepackage[top=0.75in, bottom=1in, left=0.625in, right=0.625in]{geometry}
\usepackage{fancyhdr}
\fancypagestyle{firstpage}{%
  \fancyhf{}%
  \fancyhead[R]{\scriptsize\thepage}%
  \fancyfoot[C]{\footnotesize This work has been accepted for publication in the Proceedings of the IEEE Global Communications Conference (\textit{GLOBECOM}) 2026.}%
}
\def\BibTeX{{\rm B\kern-.05em{\sc i\kern-.025em b}\kern-.08em
    T\kern-.1667em\lower.7ex\hbox{E}\kern-.125emX}}
\begin{document}
\bstctlcite{IEEEexample:BSTcontrol}

\title{Delay-Robust Deep Reinforcement Learning for Ranging-Free Channel Access under Mobility in Underwater Acoustic Networks}

\author{
    \IEEEauthorblockN{Huaisheng Ye\textsuperscript{1}, Xiaowen Ye\textsuperscript{2}, and Liqun Fu\textsuperscript{1}
    \thanks{This work was supported by the National Natural Science Foundation of China under Grant U23A20281.}%
    \thanks{\emph{Corresponding author: Liqun Fu.} The source code of this work is available at \protect\url{https://github.com/HysonYe/CHILL-STER}.}}
    
    \IEEEauthorblockA{\textsuperscript{1}School of Informatics, Xiamen University, Xiamen 361102, China}
    
    \IEEEauthorblockA{\textsuperscript{2}College of Photonic and Electronic Engineering, Fujian Normal University, Fuzhou 350117, China}
    
    \IEEEauthorblockA{Email: huaishengye@stu.xmu.edu.cn, xiaowenye@fjnu.edu.cn, liqun@xmu.edu.cn}
}

\maketitle
\thispagestyle{firstpage}

\begin{abstract}
Long propagation delays in underwater acoustic networks (UWANs) cause spatio-temporal uncertainty, constraining channel utilization in medium access control (MAC) protocols. Node mobility within autonomous underwater vehicle scenarios exacerbates these challenges by introducing dynamic propagation delays and varying spatial topologies. We present MobiU-MAC, a deep reinforcement learning (DRL)-based MAC protocol for mobile node access in UWANs that maximizes throughput via autonomous learning. MobiU-MAC incorporates CHILL-STER, a novel DRL algorithm optimized for UWANs that is both ranging-free and delay-robust. CHILL-STER employs a credit horizon-limited $\lambda$-return (CHILL-Return) mechanism to achieve stable learning under asynchronous delayed rewards, while the companion spatio-temporal experience replay (STER) mechanism addresses topological changes arising from node mobility. This work also demonstrates theoretically that DRL attains optimal policy learning equivalent to a standard Markov decision process under long propagation delays without requiring ranging. Performance evaluations indicate that MobiU-MAC outperforms existing DRL-based MAC protocols for UWANs by leveraging the maximum system delay boundary without ranging overhead, supporting the effectiveness of the proposed theory and algorithm in complex underwater dynamic environments.
\end{abstract}


\section{Introduction}
Underwater acoustic networks (UWANs) primarily utilize acoustic waves for long-range communication due to the severe attenuation of electromagnetic waves in marine environments. Given narrow spectrum and Doppler spread, TDMA is more robust than FDMA/CDMA and thus the most prevalent access method in UWANs \cite{geng2023}. However, the low propagation speed of acoustic waves (~1500 m/s) induces long propagation delays that challenge the design of medium access control (MAC) protocols. This delay introduces spatio-temporal uncertainty \cite{ye2022}, which reduces the efficiency of traditional handshake and carrier-sensing mechanisms. Consequently, intelligent access methods are required to sense the dynamic environment and facilitate adaptive decisions.

Deep reinforcement learning (DRL) has proven effective in sequential decision-making for dynamic underwater environments, with demonstrated utility in link adaptation \cite{wu2025}, routing \cite{he2025}, and AUV-assisted trajectory planning \cite{zhang2024}. This trend is now driving MAC protocol designs toward intelligence for coordinated multi-node access. Early research employed tabular Q-learning to optimize access strategies \cite{sung2019}, though this approach is limited by high-dimensional state spaces \cite{geng2023, ye2022}. With the rise of deep learning, DRL has gradually become a core tool for designing intelligent MAC protocols in UWANs. For instance, the DR-DLMA protocol incorporates a DR-DQN algorithm using propagation delay information for experience replay with slot alignment, addressing the problem of training failure caused by ACK feedback delay \cite{ye2022}. The DL-MAC protocol further explores asynchronous transmission, which reduces collisions caused by spatio-temporal uncertainty through flexibly adjusting transmission start times, thereby improving network throughput \cite{geng2023}. Furthermore, multi-agent reinforcement learning (MARL)-based MAC protocols for UWANs have recently emerged \cite{gao2024, huang2025}.

However, existing DRL-based methods involve a trade-off regarding the decision interval. In time-division MAC protocols, the slot duration is often directly coupled with the decision time-step, making this contradiction especially prominent. Existing research falls into conservative and radical schemes. Conservative schemes use a time-step exceeding the maximum one-way propagation delay \cite{geng2023}, avoiding delayed feedback but reducing time-domain utilization. Radical schemes use a time-step much smaller than the maximum one-way propagation delay \cite{ye2022, gao2024, huang2025}, maximizing time-domain utilization and network performance at the cost of severe delays in both observation feedback and reward feedback. In dynamic topologies, reward feedback delay is particularly challenging and remains unresolved. In such networks, ranging-based reward alignment is impractical due to high overhead, leading to a credit assignment problem that hinders the convergence of DRL agents.

To overcome these limitations, this paper adopts the radical scheme and proposes CHILL-STER, a novel DRL algorithm tailored for the underwater acoustic environments. Its core advantages are that it is ranging-free and exhibits delay-robust. The algorithm requires no real-time ranging overhead and enables efficient learning using only the maximum delay bound of the system. Building on this, this paper develops MobiU-MAC, an intelligent MAC protocol designed for mobile node access scenarios. The protocol can adaptively sense topology changes and significantly improve network throughput in dynamic environments through optimized access policies.
\begin{enumerate}[label=\arabic*)]
    \item For the first time, we prove that learning the optimal policy in underwater environments with long propagation delays is equivalent to that in a standard Markov decision process (MDP). Specifically, by revealing the intrinsic mathematical relationship between delay bounds and return estimation in UWANs, we show that DRL algorithms based on multi-step returns (e.g., n-step return, $\lambda$-return \cite{sutton1998}) retain convergence to the optimal strategy of a standard MDP, even without explicit delay compensation.
    \item We introduce \textbf{CHILL-STER}, a ranging-free and delay-robust algorithm comprising a \textit{Credit Horizon-Limited $\lambda$-return} (CHILL-Return) to enable convergence under asynchronous rewards and a \textit{Spatio-Temporal Experience Replay} (STER) to mitigate node mobility-induced instability.
    \item We develop and evaluate \textbf{MobiU-MAC}, an intelligent MAC protocol for mobile node access. Simulation results indicate that MobiU-MAC improves channel utilization and robustness in high-speed mobility scenarios.
\end{enumerate}

\section{System Model}
\begin{figure}[!t]
    \centering
    \includegraphics[width=0.92\columnwidth, trim=0 0 0 0, clip]{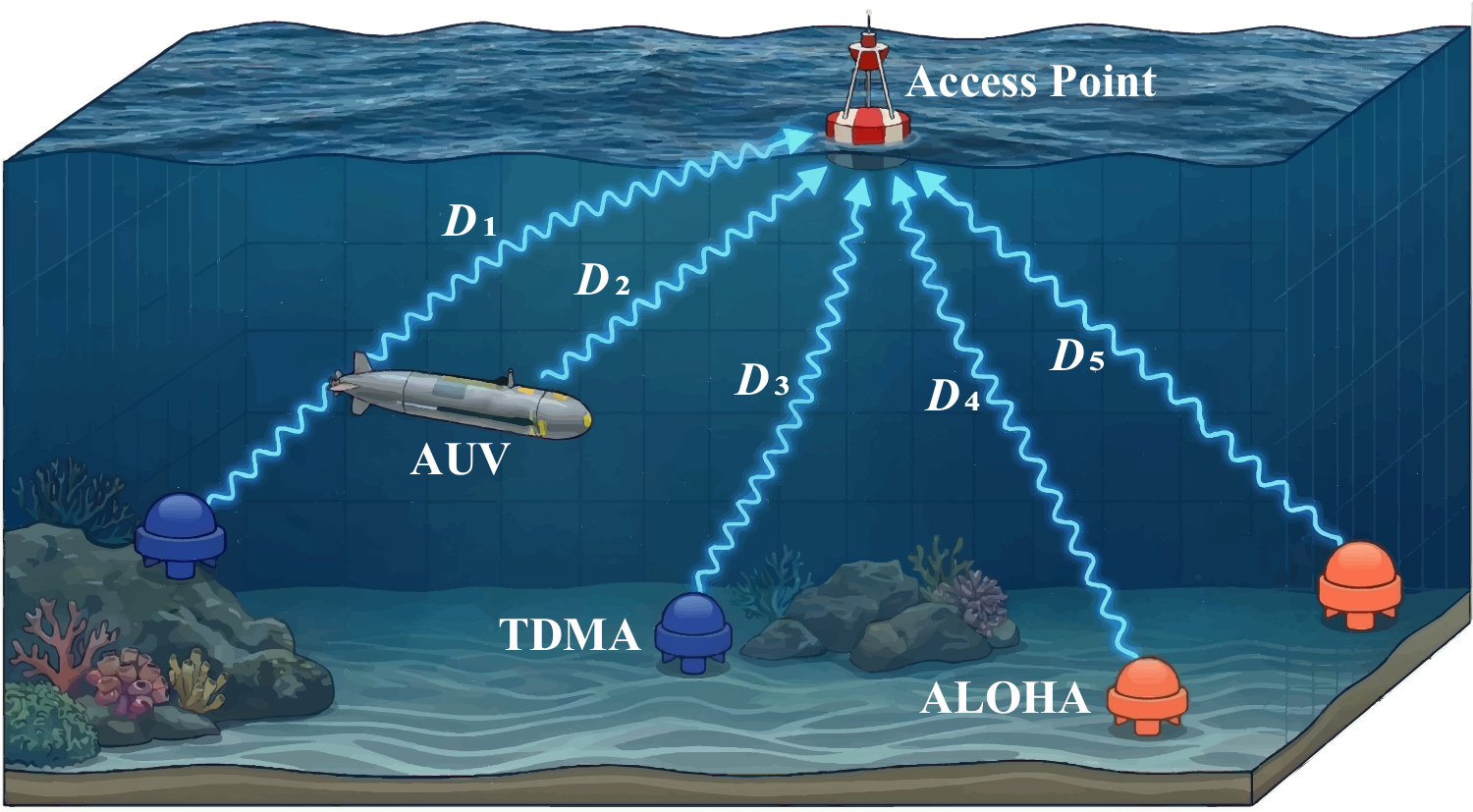}
    \caption{System model of the heterogeneous UWANs.}
    \label{fig:system model}
    \vspace{-1.5em}
\end{figure}

We consider a heterogeneous UWAN with an Access Point (AP) and $N$ underwater nodes (one mobile AUV and $N-1$ static nodes) sharing an uplink channel, as illustrated in Fig.~\ref{fig:system model}. The AUV, moving continuously to perform underwater operations, is equipped with a DRL-based MAC protocol and acts as a DRL agent. Concurrently, all underwater nodes continuously upload traffic data to the AP. To emulate a heterogeneous environment, static nodes each follow either TDMA (pre-allocated slots) or ALOHA (transmission probability $q$).

Following the setup in \cite{ye2022}, the system is time-slotted with duration $\tau$, where each slot covers one uniform-length data packet and its corresponding ACK. In each time slot, a node decides to either transmit or wait according to its MAC protocol. The AP listens for data in the first part of each time slot and can successfully receive at most one packet. Multiple packets arriving in the same time slot cause a collision. Otherwise, the AP receives a data packet from a node, it broadcasts an ACK packet containing the MAC address of that node in the latter part of the time slot.

As underwater acoustic propagation is far slower than radio waves, propagation delay is significant in UWANs. The one-way propagation delay for node $i$ at distance $d_i$ is $D_i$ slots. Specifically,
\begin{equation}
D_i = \left\lceil \frac{d_i}{c \tau} \right\rceil,
\label{eq:Di}
\end{equation}
where $c$ is the underwater sound speed. As the AUV moves, its distance to the AP and the propagation delay $D(t)$ vary, causing a time-varying topology and enhanced spatio-temporal uncertainty. The agent must then autonomously learn to access time slots not occupied by static nodes during AUV movement, maximizing total network throughput.

\section{Deep Reinforcement Learning for UWANs with Delayed Rewards}
\subsection{Markov Decision Process and Preliminaries of DRL}
Standard Markov decision processes are defined by the tuple $(\mathcal{S}, \mathcal{A}, P, R, \gamma)$, where $\mathcal{S}$ and $\mathcal{A}$ are finite sets of states and actions, $P: \mathcal{S} \times \mathcal{A} \times \mathcal{S} \rightarrow [0,1]$ satisfies $P(s'|s,a) = \Pr(s_{t+1}=s' | s_t=s, a_t=a)$, while $R: \mathcal{S} \times \mathcal{A} \rightarrow \mathbb{R}$ denotes the immediate reward function and $\gamma \in [0,1)$ is the discount factor \cite{sutton1998}. At step $t$, the agent executes $a_t \sim \pi(s_t)$, receives $r_{t+1}$, and transitions to $s_{t+1}$. The agent aims to find an optimal policy $\pi^*$ maximizing the expected return $G_t = \mathbb{E}_\pi [ \sum_{k=0}^{\infty} \gamma^k r_{t+k+1} ]$.

The optimal action-value function $Q^*(s, a)$ satisfies the Bellman equation $Q^*(s, a) = \mathbb{E} [ R(s, a) + \gamma \max_{a'} Q^*(s', a') ]$. To handle large or continuous state spaces, the $Q$-function is approximated using a deep neural network $Q(s,a;\theta)$ with parameters $\theta$. The deep $Q$-network (DQN) algorithm \cite{mnih2015human} minimizes the temporal-difference (TD) loss:
\begin{equation}
\mathcal{L}(\theta) = \mathbb{E}_{(s,a,r,s') \sim \mathcal{D}} \left[ \left( Q(s,a;\theta) - y \right)^2 \right],
\label{eq:td_loss}
\end{equation}
where the target value is $y = r + \gamma \max_{a'} Q(s',a';\theta^-)$, with $\mathcal{D}$ the experience replay buffer storing past transitions and $\theta^-$ the target network parameters updated periodically to stabilize training.

However, this standard formulation assumes that the reward $r$ is observed immediately after taking action $a$, an assumption that is violated in UWANs due to long propagation delays.

\subsection{Challenges of Standard DRL: Reward Feedback Delay and Asynchrony}
While observation delay is addressed by incorporating historical observations into the state representation \cite{ye2022, geng2023}, this paper focuses on the more critical \textbf{reward delay} problem. This delay stems from the round-trip propagation of ACKs, which violates the standard DRL assumption.

\begin{figure}[!t]
    \centering
    \includegraphics[width=0.9\columnwidth, trim=0 0 0 0, clip]{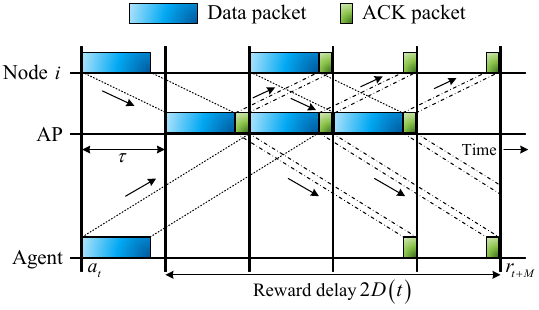}
    \vspace{-1em}
    \caption{Impact of long propagation delay on reward observation. Due to the round-trip propagation delay $2D(t)$ , the reward $r_{t+M}$ for action $a_t$ is only observed at time step $t+M$ , where $M=2D(t)+1$ .}
    \label{fig:delay_illustration}
    \vspace{-1.8em}
\end{figure}

Fig.~\ref{fig:delay_illustration} illustrates that under the radical scheme, the reward for $a_t$ is only observed at time step $t+M$, where $M = 2D(t) +~1$. Crucially, rewards in the window $[t+1, t+M-1]$ are independent of at and depend instead on the environment or other nodes. Single-step TD updates incorrectly attribute this noise to $a_t$, causing non-convergence.

Existing DRL methods \cite{ye2022} align rewards using ranging-based estimates of $D(t)$. Upon receiving the complete ACK at the end of time slot $t + 2D(t)$, the agent constructs $(s_t, a_t, r_{t+2D(t)+1}, s_{t+1})$ for standard DRL updates, restoring the assumption that reward follows immediately after the action. However, ranging incurs high overhead and is impractical for mobile nodes with fast-changing topologies.

\subsection{Theoretical Analysis: Ranging-Free Optimal Policy Equivalence Under Delays}
\subsubsection{Optimal Policy Equivalence}
We assume a node with a fixed one-way propagation delay $D$ described in \eqref{eq:Di}. The reward for action $a_t$ is observed following a round-trip delay $M = 2D + 1$. The aligned Q-function for a standard MDP with immediate rewards is defined as $Q_{\text{align}}(s_t, a_t) = \mathbb{E}_{\pi} \left[ \sum_{n=0}^{\infty} \gamma^{n} r_{t+M+n} \;\middle|\; s_t, a_t \right]$.

In the actual delayed environment, the return associated with $a_t$ includes $M-1$ irrelevant rewards:
\[
Q(s_t, a_t) = \mathbb{E}_{\pi} [ \sum_{m=0}^{M-2} \gamma^m r_{t+m+1} + \sum_{n=0}^{\infty} \gamma^{M+n-1} r_{t+M+n} ]_{s_t, a_t}
\]
The first $M-1$ rewards $\{r_{t+1},\dots,r_{t+M-1}\}$ are independent of $a_t$; they depend only on state $s_t$. Denote this state-dependent bias as $V(s_t) = \mathbb{E}_{\pi} \left[ \sum_{m=0}^{M-2} \gamma^{m} r_{t+m+1} \;\middle|\; s_t \right]$. Thus $Q$ decomposes as $Q(s_t, a_t) = V(s_t) + \gamma^{M-1} Q_{\text{align}}(s_t, a_t)$. Since $\gamma^{M-1}>0$ and $V(s_t)$ does not vary with $a_t$, for any fixed $s_t$:
\begin{equation}
\arg\max_{a} Q(s_t, a) = \arg\max_{a} Q_{\text{align}}(s_t, a).
\end{equation}
This equivalence directly implies the following theorem.

\textbf{Theorem 1 (Policy Equivalence)}: Maximizing the unaligned return without ranging or reward alignment yields the same optimal policy as the aligned standard MDP.

\subsubsection{Multi-Step Return Horizon}
Although Theorem 1 guarantees equivalence, single-step TD learning fails because its TD target $y_{\text{1step}} = r_{t+1} + \gamma \max_{a'} Q(s_{t+1}, a')$ (see the target $y$ in \eqref{eq:td_loss}) uses $r_{t+1}$, which is uncorrelated with $a_t$, leading to high-variance gradients.

Reliable $Q$ estimation requires a multi-step return horizon $H = M$ spanning the entire delay window, yielding the following multi-step return target:
\[
y_{\text{nstep}} = \underbrace{\sum_{m=0}^{M-2} \gamma^m r_{t+m+1}}_{V_{\text{noise}}} + \gamma^{M-1} r_{t+M} + {\gamma^M \max_{a'} Q(s_{t+M}, a')}, \hspace{-1.5pt}
\]
where $V_{\text{noise}}$ is the action-independent noise and $r_{t+M}$ is the true feedback. Although $V_{\text{noise}}$ adds variance to the gradient, the presence of $\gamma^{M-1} r_{t+M}$ provides a genuine signal that guides the update toward the optimal policy.

For mobile nodes, we utilize a prior delay bound $D_{\max} = \lceil d_{\max}/(c_{\min} \cdot \tau) \rceil$ based on maximum distance. This yields the following design criterion.

\textbf{Design Criterion}: Establishing the mathematical relationship between the multi-step return horizon $H$ and the delay bound,
\begin{equation}
H \ge 2D_{\max} + 1, \label{eq:multi_step_horizon}
\end{equation}
ensures that the return contains the reward signal $r_{t+2D(t)+1}$ for any delay $D(t) \le D_{\max}$. This enables stable convergence to the optimal policy without any real-time ranging overhead.

Unlike \cite{su2026, paulo2022} which use multi-step returns empirically, this work establishes the theoretical policy equivalence under long delays.

\section{MobiU-MAC: A Delay-Robust and Ranging-Free MAC Protocol Driven by CHILL-STER}

\subsection{CHILL-STER Algorithm: Dual Mechanisms for Underwater Challenges}
Although Section~III theoretically guarantees policy optimality with a multi-step return horizon (see \eqref{eq:multi_step_horizon}), naive multi-step methods suffer from high variance and on-policy bias, while node mobility further introduces non-stationary topology dynamics. To address these challenges, we propose \textbf{CHILL-STER}, integrating \textit{Credit Horizon-Limited \(\lambda\)-return} (CHILL-Return) for delay-robustness and \textit{Spatio-Temporal Experience Replay} (STER) for robustness to topology changes.

\subsubsection{Credit Horizon-Limited $\lambda$-Return}

CHILL-Return balances the bias-variance trade-off by restricting the $\lambda$-return to a compact horizon $\hat{H}$ (satisfying \eqref{eq:multi_step_horizon}). Off-policy bias is mitigated via importance sampling.

The $\hat{H}$-step return is
\begin{equation}
G_t^{\lambda,\hat{H}} = \max_{a} Q(s_{t}, a) + \sum_{k=0}^{\hat{H}-1} (\gamma\lambda)^k \delta_{t+k},
\label{eq:chill_return}
\end{equation}
where $\delta_t = r_{t+1} + \gamma \max_{a'} Q(s_{t+1}, a') - \max_{a} Q(s_t, a)$ is the TD error. We apply a clipped importance weight
\begin{equation}
w_t = \min\left(1, \; \rho_{t+1:t+\hat{H}-1}^{\beta}\right), \quad \beta \in (0,1],
\label{eq:weight_clip}
\end{equation}
based on trajectory-wise importance factor $\rho_{t+1:t+\hat{H}-1} = \prod_{k=1}^{\hat{H}-1} \frac{\pi(a_{t+k} \mid s_{t+k})}{\mu(a_{t+k} \mid s_{t+k})}$, where $\pi$ is the target policy and $\mu$ is the behavior policy. The hyperparameter $\beta \in (0,1]$ controls the degree of correction applied.

Combining the $\hat{H}$-step return with importance sampling, CHILL-Return minimizes the following loss to update the value network:
\begin{equation}
\mathcal{L}(\theta) = \mathbb{E}_{(s_t, a_t) \sim \mathcal{D}} \left[ w_t \left( G_t^{\lambda,\hat{H}} - Q(s_t, a_t; \theta) \right)^2 \right].
\label{eq:chill_loss}
\end{equation}
While we present CHILL-Return in the action-value form $Q(s,a)$ for concreteness, the same credit horizon-limiting principles can be extended to other value-based frameworks (e.g., DDPG, or state-value $V(s)$ with suitable bootstrapping).

\subsubsection{Spatio-Temporal Experience Replay}
AUV mobility introduces non-stationarity that traditional temporal-only experience replay (ER) fails to capture.

We propose STER, which assigns an implicit spatial anchor $z_t \in \mathcal{Z} \subseteq \mathbb{R}$ (e.g., ranging, dead reckoning, or signal feature estimation) to each sample and stores transitions in independent buffers $\mathcal{D}_z$ based on the node's relative position. To stabilize sampling, a smoothed spatial context $c_t$ is updated via $c_t = \alpha c_{t-1} + (1-\alpha) z_t$, where $\alpha \in [0,1)$ is the smoothing factor. The smoothed context reflects the long-term spatial trend of the node.

For a context $c$ and radius $R$, we sample from neighboring anchors $\mathcal{Z}_{\text{near}}(c) = \{ z \in \mathcal{Z}_{\text{visited}} : |z-c| \le R \}$. The sampling quota $b(z)$ for each anchor $z$ follows a distance-decay principle: $b(z) \propto w(z) = f(|z-c|)$, where $f(\cdot)$ is a monotonically decreasing function (e.g., $f(d)=2^{-d}$). Quotas are normalized to sum to batch size $B$, and $b(z)$ segments of length $\hat{H}$ are sampled from each $\mathcal{D}_z$ to form the training batch, effectively mitigating boundary effects.

\subsection{MobiU-MAC Protocol Design: Integrating CHILL-STER for Channel Access under Mobility}

Algorithm~\ref{alg:mobiu-mac} details the joint operation of CHILL-Return and STER.

\subsubsection{MDP Formulation}
We model the MAC interaction as a Partially Observable MDP (POMDP) with a decision interval $\tau$ matching the slot duration.

\begin{itemize}[leftmargin=*]
    \item \textbf{Action Space:} The agent selects $a_t \in \{a_{\mathrm{wait}}, a_{\mathrm{tx}}\}$ per slot $ t $, representing silent or transmission modes.
    
    \item \textbf{State Space:} Given unobservable global states, the agent uses local observations $o_t = (a_{t-1}, \xi_{t-1})$, where $\xi_{t-1} \in \{\xi_{\mathrm{fail}}, \xi_{\mathrm{succ}}, \xi_{\mathrm{busy}}\}$ represents the ACK feedback received at the end of the previous slot. The variables $\xi_{\mathrm{fail}}$, $\xi_{\mathrm{succ}}$, and $\xi_{\mathrm{busy}}$ denote an idle/collision slot, a successful self-transmission, and a successful transmission by another node, respectively. To mitigate observation delays, the agent constructs $s_t$ by stacking the $L_{\text{hist}}$ most recent observations $s_t = (o_{t-L_{\text{hist}}+1}, o_{t-L_{\text{hist}}+2}, \dots, o_t) \in \mathcal{S}$.

    \item \textbf{Reward Function:} To maximize utilization, $r_{t+1} = 1$ if $\xi_t \in \{\xi_{\mathrm{succ}}, \xi_{\mathrm{busy}}\}$ and $0$ otherwise, incentivizing successful access or yielding to other nodes.
\end{itemize}
\subsubsection{Implicit Spatial Anchor Estimation}
To support STER without ranging overhead, an implicit spatial anchor $z_t$ is estimated by exploiting the statistical correlation between actions and ACK feedback. The core principle holds that the propagation delay $D(t)$ dictates the temporal coupling between a transmission $a_t$ and its corresponding ACK $\xi_{t+2D(t)}$. For candidate offsets $m \in \{2, \dots, 2D_{\max}\}$, we maintain a cumulative score $S(m) = \sum_{k=t-W}^{t-m} \phi(a_k, \xi_{k+m})$ over a sliding window $W$. The scoring function $\phi(a, \xi)$ is 1 for alignment ($(a_{\text{tx}}, \xi_{\text{succ}})$ or $(a_{\text{wait}}, \xi_{\text{fail}})$), -1 for misalignment ($(a_{\text{tx}}, \xi_{\text{busy}})$ or $(a_{\text{wait}}, \xi_{\text{succ}})$), and 0 otherwise.

The optimal offset $\hat{m}_t = \arg\max_{m} S(m)$ is smoothed via a moving average to filter instantaneous noise. Finally, the spatial anchor is derived as $z_t = \lceil m_t / 2 \rceil$. This mechanism enables the agent to categorize experiences into spatial regions based on inherent communication patterns without active signaling overhead.

\begin{algorithm}[t]
\caption{MobiU-MAC Protocol}
\label{alg:mobiu-mac}
\begin{algorithmic}[1]
\setlength{\baselineskip}{11pt}
\REQUIRE Maximum delay bound $D_{\max}$;

        hyperparameters $\hat{H}$, $\gamma$, $\lambda$, $\beta$, $\alpha$, $B$, $L_{\text{hist}}$, $R$
\STATE Initialize $Q_\theta$, $Q_{\theta^-}$, STER buffer $\{\mathcal{D}_z\}$, spatial context $c\leftarrow0$, history buffer $\mathcal{O}$ (length $L_{\text{hist}}$), initial state $s_0$
\FOR{each time slot $t$}
    \STATE Select action $a_t \sim \varepsilon\text{-greedy}(Q_\theta(s_t,\cdot))$
    \STATE Execute $a_t$ (i.e., transmit or wait), receive observation $o_{t+1} = (a_t, \xi_t)$, then compute reward $r_{t+1}$ from $\xi_t$
    \STATE Update $\mathcal{O} \leftarrow \mathcal{O} \cup \{o_{t+1}\}$, construct state $s_{t+1}$ from $\mathcal{O}$
    \STATE $\blacktriangleright$ \textbf{STER:}
    \STATE Estimate implicit spatial anchor $z_t$
    \STATE Update spatial context $c \leftarrow \alpha c + (1-\alpha) z_t$
    \STATE Store transition $(s_t, a_t, r_{t+1}, s_{t+1})$ into $\mathcal{D}_{z_t}$
    \STATE Sample $\mathcal{B}$ from $\bigcup\limits_{z: |z-c| \le R} \mathcal{D}_z$ with distance-decay quotas
    \STATE $\blacktriangleright$ \textbf{CHILL-Return:}
    \FOR{each sample $i$, take trajectory length $\hat{H}$ from $i$}
        \STATE Compute TD errors $\delta_{i+k}$, $k=0,\dots,\hat{H}-1$
        \STATE Compute $\hat{H}$-step return $\displaystyle G_i^{\lambda,\hat{H}}$
        \STATE Compute importance factor $\rho = \prod_{k=1}^{H-1} \frac{\pi(a_{i+k}|s_{i+k})}{\mu(a_{i+k}|s_{i+k})}$
        \STATE Compute importance weight $w_i = \min\!\big(1, \rho^{\beta}\big)$
    \ENDFOR
    \STATE Update $\theta$ by minimizing $\frac{1}{|\mathcal{B}|} \sum w_i \big(G_i^{\lambda,H} - Q_\theta(s_i,a_i)\big)^2$
    \STATE Periodically update $\theta^- \leftarrow \theta$
\ENDFOR
\end{algorithmic}
\end{algorithm}
\vspace{-0.6em}
\section{Evaluations}
\begin{figure*}[!t]
    \centering
    \includegraphics[width=0.99\textwidth]{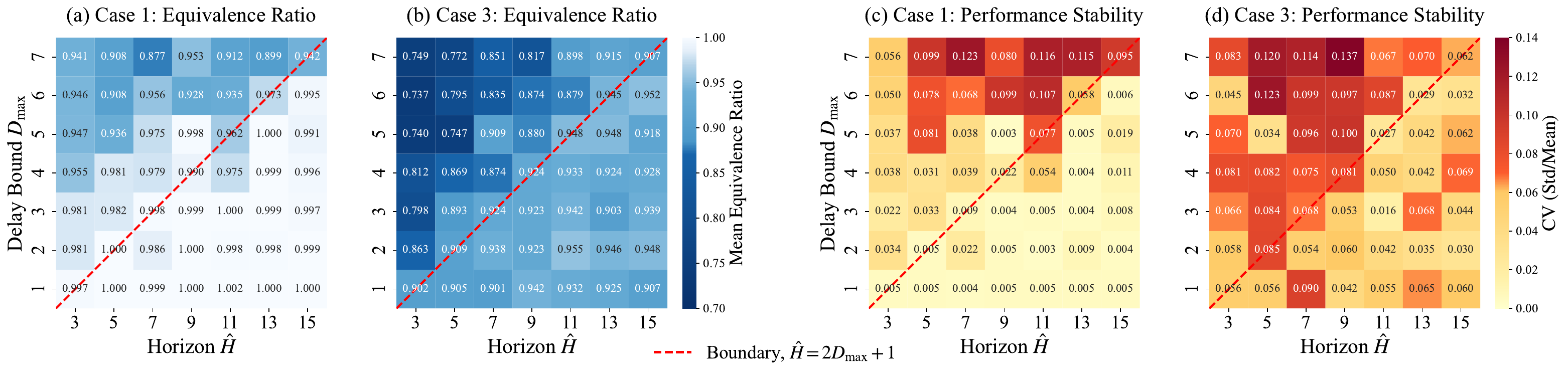}
    \vspace{-1em}
    \caption{Mechanistic analysis of equivalence ratio and performance stability across varying $D_{\max}$ and $\hat{H}$. The red dashed line represents the theoretical boundary $ \hat{H} = 2D_{\max} + 1 $, beyond which the MobiU-MAC achieves near-optimal performance equivalent to a standard MDP as established in Theorem 1.}
    \label{fig:eval_horizon}
    \vspace{-1.5em}
\end{figure*}
\subsection{Simulation Settings and Parameters}
\vspace{-0.3em}
Simulations are conducted in a 3D underwater acoustic network of size $1000 \times 1000 \times 100$ m$^3$. The AP is located at the sea-surface center $(0, 0, 0)$ with sound speed $c = 1500$ m/s, slot duration $\tau = 0.1$ s, and $D_{\max} = 5$ slots. We evaluate four scenarios: \textbf{Case~1} includes a mobile AUV coexisting with four TDMA nodes where three nodes occupy one fixed slot each and one node occupies two slots within a 10-slot frame; \textbf{Case~2} features a mobile AUV and an ALOHA node ($q=0.2$); \textbf{Case~3} combines Cases~1 and~2; and \textbf{Case~4} replicates Case~1 but introduces a random slot reallocation for the two-slot TDMA node at $t=2.5\times 10^4$ slots to evaluate robustness.

The AUV runs the proposed MobiU-MAC protocol to learn access policies autonomously. In stationary conditions, the AUV is located at $(480, 480, 10)$. In mobile settings, the AUV moves at constant velocities $v \in \{1, 2, 6, 10, 15, 20, 30\}$ m/s along random waypoints. We evaluated the system in an online learning mode over $5 \times 10^4$ slots.

MobiU-MAC parameters include: $\gamma = 0.978$ , $\lambda = 0.971$ , $\hat{H} = 12$ , $\beta = 0.2$ , $L_{\text{hist}} = 30$ , $B = 128$ , $lr = 6 \times 10^{-4}$ , $R = 2$ , $\alpha = 0.95$ , $W = 100$ , and $|\mathcal{D}_z| = 2048$. The target network updates every 60 slots, with an anchor-specific $\varepsilon$-greedy exploration decaying from $1.0$ to $0.01$ at a factor of $0.996$ for each spatial domain independently. All DRL-based methods (MobiU-MAC and benchmarks) utilize the same lightweight network architecture as in \cite{ye2022}, consisting of six linear layers with 64 neurons each.

\subsection{Benchmarks and Performance Metrics}

\subsubsection{Benchmarks}

MobiU-MAC is evaluated against two key DRL-based MAC baselines and relevant variants.

\textbf{DR-DLMA and its Oracle variant:} DR-DLMA~\cite{ye2022} is a radical scheme baseline for static nodes. We implement an Oracle variant with ideal real-time delay $ D(t) $ and continuous training, which enables the extension of the protocol from static nodes to mobile AUV scenarios, to establish a performance upper bound. The learning rate is tuned to $6 \times 10^{-4}$.

\textbf{DL-MAC and its fine-tuned variants:} async-DL-MAC \cite{geng2023} is reproduced with a decision cycle of 6 slots, denoted as \textit{DL-MAC-P6}. The longer decision cycle renders this method largely immune to delayed feedback. Furthermore, we provide fine-tuned versions with a learning rate of $1.2 \times 10^{-3}$ and decision cycles of 6 and 3 slots, denoted as \textit{DL-MAC-P6-FT} and \textit{DL-MAC-P3-FT}, respectively.

All baselines adopt the same network architecture and $\epsilon$-greedy exploration strategy consistent with MobiU-MAC. Hyperparameters not specified follow their original papers.

\subsubsection{Performance Metrics}

We evaluate \textbf{Running Average Throughput} $\eta_{\text{avg}}(t) = \frac{1}{L_{\text{win}}} \sum_{k=t-L_{\text{win}}+1}^{t} \mathbf{1}_{\{\xi_k \in \{\xi_{\mathrm{succ}}, \xi_{\mathrm{busy}}\}\}}$ with $L_{\text{win}} = 2000$ to track learning dynamics ($\mathbf{1}_{\{\cdot\}}$ is the indicator function); and \textbf{Steady-state Throughput} $\eta_{\text{ss}}$ averaged over the final $10^4$ slots to assess converged performance.

All experiments involved 10 independent runs. Performance curves depict the mean, where shaded regions indicate the standard deviation unless otherwise noted.

\subsection{Performance in Static Heterogeneous UWANs}

We evaluate MobiU-MAC in static heterogeneous environments by analyzing steady-state throughput and resilience to strategy shifts.

\subsubsection{Steady-state Throughput Analysis}

\begin{figure}[t]
    \centering
    \includegraphics[width=1\columnwidth, trim=0 0 0 0, clip]{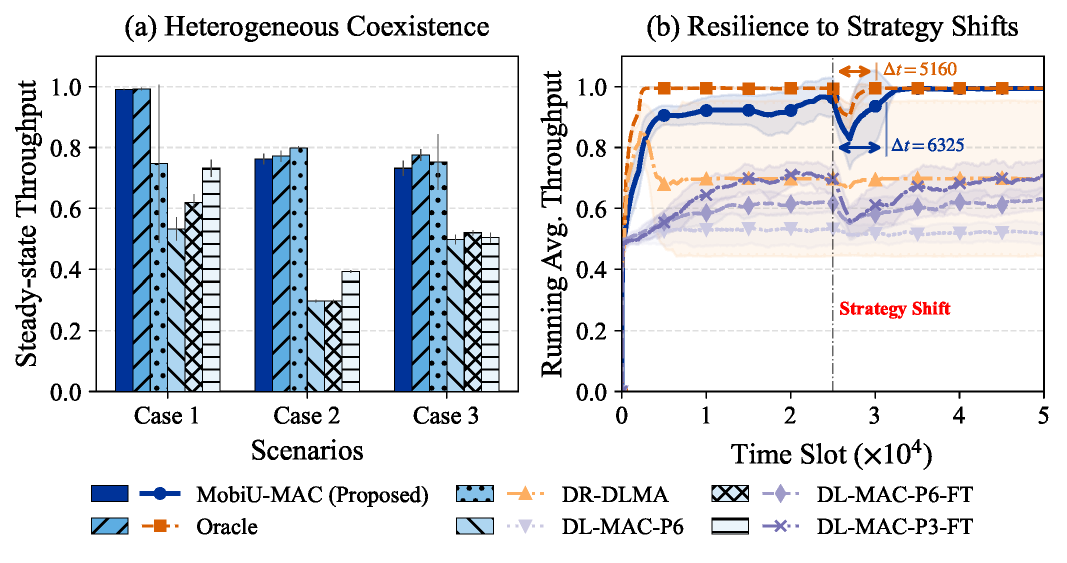}
    \vspace{-2.3em}
    \caption{Performance under heterogeneous coexistence and strategy shifts.}
    \label{fig:eval_static}
    \vspace{-1.5em}
\end{figure}

Fig.~\ref{fig:eval_static}(a) shows that MobiU-MAC achieves near-optimal steady-state throughput across Cases 1–3, with a gap to Oracle within 0.2\%–5.7\%. This validates Theorem 1, demonstrating that ranging-free learning can attain optimal policies.
Performance degrades for all protocols in Cases~2 and~3. Because ALOHA nodes lower the throughput ceiling and increase randomness, which particularly hinders DL-MAC due to its joint learning of timing and transmission decisions. This higher-dimensional action space slows convergence. Compared to DR-DLMA, MobiU-MAC exhibits lower standard deviation than DR-DLMA in Cases 1 and 3, indicating more stable convergence.

\subsubsection{Resilience to Strategy Shifts}

In Case 4, MobiU-MAC recovers from strategy shifts at $t=2.5 \times 10^4$ in $6325$ slots, comparable to Oracle's $5160$ slots. The slight delay is due to MobiU-MAC's larger replay buffer retaining stale experiences longer. Conversely, DL-MAC recovers slower due to its decision cycle limiting sample efficiency.

DR-DLMA exhibits a sharp throughput drop near the 2000-th slot due to its heuristic training suspension, which may halt learning before convergence.

\subsection{Robustness against AUV Mobility}

Under AUV mobility (Case 1), MobiU-MAC maintains steady-state throughput generally above $80\%$ across all speeds. Conversely, Oracle's performance drops below MobiU-MAC when $v > 2$ m/s despite perfect delay info, while DR-DLMA and DL-MAC suffer significant degradation. The tested velocity range extends well past typical operational AUV cruise speeds to stress-test robustness, and MobiU-MAC remains stable even at the extreme $30$ m/s.
\begin{figure}[t]
    \centering
    \includegraphics[width=1\columnwidth, trim=0 0 0 0, clip]{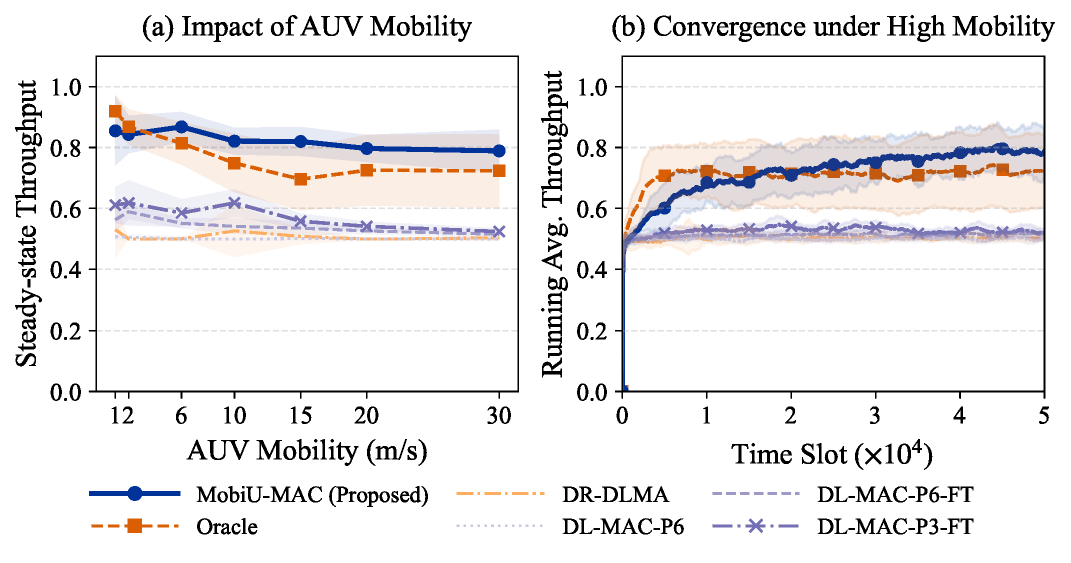}
    \vspace{-2.3em}
    \caption{Throughput performance and convergence dynamics under AUV mobility. (a) Steady-state throughput vs. AUV velocity. (b) Running avg. throughput at $v = 30$ m/s.}
    \label{fig:eval_mobility}
    \vspace{-1em}
\end{figure}

This superiority stems from CHILL-STER: CHILL-Return ensures ranging-free convergence under long delays , whereas DR-DLMA's ranging-dependency and heuristic training fail in dynamic environments. STER further captures spatial characteristics via implicit anchors, which traditional experience replay misses.

\subsection{Mechanistic Analysis and Ablation Study}

\subsubsection{Mechanistic Analysis}
We evaluate the Equivalence Ratio $\eta_{\text{ss}} / \eta^{*}$ (where $\eta^{*}$ is the Oracle throughput) to assess ranging-free policy learning. Fig.~\ref{fig:eval_horizon} confirms a threshold effect: the equivalence ratio approaches 1.0 only when $\hat{H} \ge 2D_{\max} + 1$, empirically corroborating the policy equivalence in Theorem 1. Although Case 3 exhibits lower ratios due to the non-stationarity of ALOHA nodes, the criterion remains a critical inflection point for both throughput and stability, as evidenced by the \textit{Coefficient of Variation} (CV) heatmap.

\begin{figure}[t]
    \centering
    \includegraphics[width=0.92\columnwidth, trim=0 0 0 0, clip]{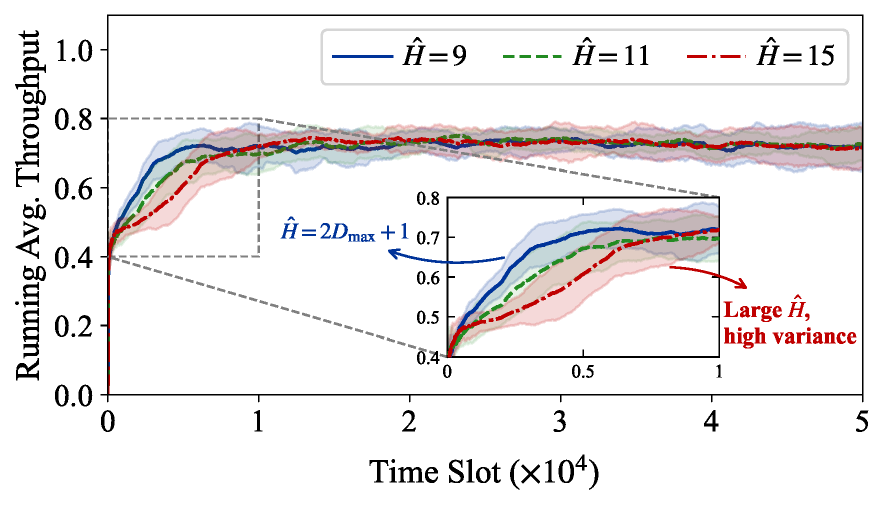}
    \vspace{-1.2em}
    \caption{Convergence dynamics of running average throughput for different horizons $\hat{H}$ in Case 3 ($D_{\max} = 4$).}
    \label{fig:show_n_step}
    \vspace{-1.6em}
\end{figure}

However, excessive $\hat{H}$ (e.g., $\hat{H}=15$ for $D_{\max}=4$) increases variance and slows convergence compared to a tighter bound $\hat{H}=9$ (Fig.~\ref{fig:show_n_step}), suggesting that $\hat{H}$ should be set slightly above the minimum integer bound.

\subsubsection{Ablation Study}

Ablations in Fig.~\ref{fig:eval_static}(a) identify multi-step return (MSR) as the primary enabler for learning under delays, as disabling it (w/o MSR) reduces performance to DQN levels. Importance sampling (w/o IS) and $\lambda$-return (w/o $\lambda$) are also essential for off-policy correction and variance reduction.
\begin{figure}[t]
    \centering
    \includegraphics[width=1\columnwidth, trim=0 0 0 0, clip]{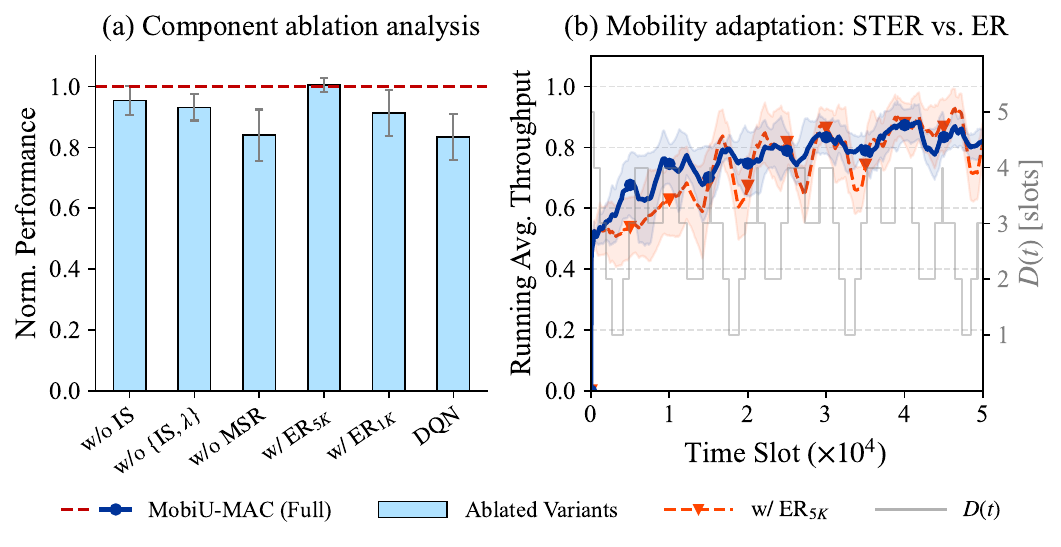}
    \vspace{-2.3em}
    \caption{Performance analysis: (a) Component ablation study with results normalized (Norm.) to MobiU-MAC (Full); (b) Comparison of mobility adaptation between STER and ER under time-varying propagation delay $ D(t) $.}
    \label{fig:show_ablation_bar}
    \vspace{-1.5em}
\end{figure}

Replacing STER with standard ER (w/ ER) fails in dynamic scenarios ; even larger buffers (e.g., $\text{ER}_{5K}, K=|\mathcal{D}_z|$ ) suffer from stale experience interference as $ D(t) $ varies (Fig.~\ref{fig:eval_static}(b)), causing sharp throughput drops. This underscores STER's efficacy in using implicit spatial anchors to filter spatially-relevant transitions for robust mobile access.

\section{Conclusions}
This paper proposed MobiU-MAC, a ranging-free DRL-based MAC protocol for mobile node access in UWANs. We proved that multi-step returns with $H \ge 2D_{\max}+1$ guarantee convergence to the standard MDP optimum under long delays, enabling optimal policy learning without ranging overhead. The integrated CHILL-STER algorithm effectively addresses asynchronous rewards and mobility-induced non-stationarity. Simulations show that MobiU-MAC significantly outperforms state-of-the-art DRL-based protocols in both static and high-speed mobile scenarios.

\bibliographystyle{IEEEtran}
\bibliography{IEEEabrv, my_references}

@STRING{IEEE_J_IOT        = "{IEEE} Internet Things J."}

@STRING{IEEE_J_MC         = "{IEEE} Trans. Mobile Comput."}

@STRING{IEEE_J_NNLS       = "{IEEE} Trans. Neural Netw. Learn. Syst."}

@STRING{IEEE_J_OE         = "{IEEE} J. Ocean. Eng."}

@IEEEtranBSTCTL{IEEEexample:BSTcontrol,
  CTLuse_article_number     = "yes",
  CTLuse_paper              = "yes",
  CTLuse_forced_etal        = "yes",
  CTLmax_names_forced_etal  = "1",
  CTLnames_show_etal        = "1",
  CTLuse_alt_spacing        = "yes",
  CTLalt_stretch_factor     = "4",
  CTLdash_repeated_names    = "yes",
  CTLname_format_string     = "{f.~}{vv~}{ll}{, jj}",
  CTLname_latex_cmd         = "",
  CTLname_url_prefix        = "[Online]. Available:"
}

@ARTICLE{geng2023,
  author={Geng, Xuan and Zheng, Yahong Rosa},
  journal=IEEE_J_NNLS, 
  title={Exploiting Propagation Delay in Underwater Acoustic Communication Networks via Deep Reinforcement Learning}, 
  year={2023},
  volume={34},
  number={12},
  pages={10626-10637},
  doi={10.1109/TNNLS.2022.3170050}
}

@ARTICLE{ye2022,
  author={Ye, Xiaowen and Yu, Yiding and Fu, Liqun},
  journal=IEEE_J_MC, 
  title={Deep Reinforcement Learning Based MAC Protocol for Underwater Acoustic Networks}, 
  year={2022},
  volume={21},
  number={5},
  pages={1625-1638},
  doi={10.1109/TMC.2020.3029844}
}

@ARTICLE{wu2025,
  author={Shuangshuang, Wu and Chitre, Mandar and Anjangi, Prasad},
  journal=IEEE_J_OE, 
  title={Adaptive Modulation and Coding With Feedback Scheduling for an Underwater Acoustic Link}, 
  year={2025},
  volume={50},
  number={4},
  pages={3054-3073},
  doi={10.1109/JOE.2025.3585657}
}

@ARTICLE{he2025,
  author={He, Yu and Han, Guangjie and Jiang, Jinfang and Cheng, Xin and Xu, Pengfei},
  journal=IEEE_J_MC, 
  title={CADTR: Context-Aware Trust Routing Algorithm Based on Priority Sampling DDPG for UASNs}, 
  year={2025},
  volume={24},
  number={11},
  pages={11688-11702},
  doi={10.1109/TMC.2025.3581512}
}

@article{zhang2024,
    title = {A path planning method based on deep reinforcement learning for AUV in complex marine environment},
    journal = {Ocean Eng.},
    volume = {313},
    pages = {119354},
    year = {2024},
    issn = {0029-8018},
    doi = {https://doi.org/10.1016/j.oceaneng.2024.119354},
    author = {An Zhang and Weixiang Wang and Wenhao Bi and Zhanjun Huang}
}

@ARTICLE{sung2019,
  author={Park, Sung Hyun and Mitchell, Paul Daniel and Grace, David},
  journal={IEEE Access}, 
  title={Reinforcement Learning Based MAC Protocol (UW-ALOHA-Q) for Underwater Acoustic Sensor Networks}, 
  year={2019},
  volume={7},
  number={},
  pages={165531-165542},
  doi={10.1109/ACCESS.2019.2953801}
}

@inproceedings{gao2024,
    author = {Gao, Yu and Bi, Zhicheng and Wang, Chaofeng},
    title = {A Multi-Agent Reinforcement Learning-Based Transmission Protocol for Underwater Acoustic Networks},
    year = {2024},
    isbn = {9798400716744},
    publisher = {ACM},
    address = {NY, USA},
    doi = {10.1145/3631726.3631769},
    booktitle = {Proc. 17th Int. Conf. Underwater Netw. Syst.},
    articleno = {38},
    numpages = {2},
    location = {Shenzhen, China},
    series = {WUWNet '23}
}

@ARTICLE{huang2025,
  author={Huang, Jiajie and Ye, Xiaowen and Wang, Yizhe and Fu, Liqun},
  journal=IEEE_J_IOT, 
  title={Leveraging Propagation Delays: A Delay-Aware Multiagent Reinforcement Learning MAC Protocol for Underwater Acoustic Networks}, 
  year={2025},
  volume={12},
  number={20},
  pages={},
  doi={10.1109/JIOT.2025.3595133}
}

@book{sutton1998,
  title={Reinforcement learning: An introduction},
  author={Sutton, Richard S and Barto, Andrew G and others},
  volume={1},
  number={1},
  year={1998},
  publisher={MIT press Cambridge}
}

@article{su2026,
    title = {AUV path planning in complex 3D underwater environments based on an improved TD3 algorithm},
    journal = {Ocean Eng.},
    volume = {345},
    pages = {},
    year = {2026},
    issn = {0029-8018},
    doi = {https://doi.org/10.1016/j.oceaneng.2025.123688},
    author = {Chengcheng Su and Yonglong Li and Haoran Wang and Yi Xia and Jialong Li and Gang Wan and Yongcan Chen}
}

@INPROCEEDINGS{paulo2022,
  author={Padrao, Paulo and Dominguez, Alberto and Bobadilla, Leonardo and Smith, Ryan N.},
  booktitle={OCEANS}, 
  title={Towards Learning Ocean Models for Long-term Navigation in Dynamic Environments}, 
  year={2022},
  volume={},
  number={},
  pages={1-6},
  doi={10.1109/OCEANSChennai45887.2022.9775460}
}

@article{mnih2015human,
  title={Human-level control through deep reinforcement learning},
  author={Mnih, Volodymyr and Kavukcuoglu, Koray and Silver, David and Rusu, Andrei A and Veness, Joel and Bellemare, Marc G and Graves, Alex and Riedmiller, Martin and Fidjeland, Andreas K and Ostrovski, Georg and others},
  journal={nature},
  volume={518},
  number={7540},
  pages={529--533},
  year={2015},
  publisher={Nature Publishing Group}
}

\end{document}